\documentstyle[11pt,aaspp4]{article}	% preprint style
\slugcomment{Submitted to the {\it P. A. S. P.}}

\lefthead{Lauer}
\righthead{Combining Undersampled Dithered Images}

\def\shah{\rm III}
\begin{document}
\title{Combining Undersampled Dithered Images}

\author{Tod R. Lauer}
\affil{National Optical Astronomy
Observatories\altaffilmark{1}, P.~O. Box 26732, Tucson, AZ 85726}
\affil{Electronic mail: lauer@noao.edu}

\altaffiltext{1}{The National Optical Astronomy Observatories are
operated by the Association of Universities for Research in Astronomy, Inc.,
under cooperative agreement with the National Science Foundation.}

\begin{abstract}

Undersampled images, such as those produced by the {\it HST} WFPC-2,
misrepresent fine-scale structure intrinsic to the astronomical sources
being imaged.
Analyzing such images is difficult on scales close to their resolution limits
and may produce erroneous results.
A set of ``dithered'' images of an astronomical source
generally contains more information about its structure than any single
undersampled image, however, and may permit reconstruction
of a ``superimage'' with Nyquist sampling.  I present a tutorial
on a method of image reconstruction that builds a superimage
from a complex linear combination of the Fourier transforms of
a set of undersampled dithered images.  This method works by
algebraically eliminating the high order satellites in the periodic
transforms of the aliased images.  The reconstructed image is an
exact representation of the data-set with no loss of resolution
at the Nyquist scale.  The algorithm is directly
derived from the theoretical properties of aliased images and
involves no arbitrary parameters,
requiring only that the dithers are purely translational and constant
in pixel-space over the domain of the object of interest.
I show examples of its application to WFC and PC images.
I argue for its use when the best recovery of point sources or
morphological information at the {\it HST} diffraction limit is of interest.

\end{abstract}


\section{Introduction}

It's nice to work with well-sampled astronomical images.
A well-sampled image can be readily resampled to various scales,
orientations, or more complex geometries without loss of information.
Its spatial resolution is well-understood, permitting
a clear analysis of the relative contributions of information and noise.
Further, many image processing algorithms will only work on well-sampled data.
In some cases, however, it's not practical or even desirable to
obtain well-sampled images.
Given detectors with a finite number of pixels and significant
readout noise, one may prefer to trade-off resolution for
increased field size or photometric sensitivity.
Both considerations were central to the design of the {\it HST}
WFPC-1 and WFPC-2 cameras, to give examples of
instruments that produce undersampled astronomical images.
WFPC-2 in particular has generated the largest library of high-resolution
optical astronomical images to date, but ironically the severe undersampling
in the WFC system, and the still less than critical sampling of
the PC at all but the reddest wavelengths, limit the resolution
of {\it HST} observations as much as the telescope optics, themselves.

There is no magic that can undo the undersampling in a single image;
analysis of such data always requires respect for their peculiarities.
At the same time, it may be possible
to obtain additional observations with the same camera system
that contain information lost in the original images.
For example, if the camera can be offset by a fraction
of a pixel over a sequence of exposures or ``dithered,''
one can observe how the structure of objects in the image varies
with respect to their positions on the pixel-grid, and thus
recover details not contained in any single image.
This suggests that one might construct
a well-sampled super-image from a set of undersampled, but dithered images.

In general, when the size of a pixel is important with respect
to the intrinsic point-spread function (PSF), the image as observed is
\begin{equation}
I(x,y)=O(x,y)\ast P(x,y) \ast \Pi(x,y),
\end{equation}
where $O$ is the intrinsic projected appearance of the astronomical
field being imaged, $P$ is the PSF due to the telescope and camera
optics, and $\Pi$ is the spatial form of the pixel itself,
(which is often assumed to be a uniform square, although this need
not be the case), and $\ast$ means convolution.
Both $P$ and $\Pi$ limit the resolution of $I$ and thus implicitly specify
the minimum sampling requirements --- a dilemma if $\Pi$ is too
big, since it sets what the sampling really is, regardless of what's needed.
If the astronomical scene and camera are time-stable, however,
dithering the camera allows proper sampling of the field
{\it convolved with the pixel response} as well as the PSF, to be obtained.
If the camera is pointed on a fine and regular $n\times n$ grid
of sub-pixel steps, where $n$ is the number of substeps within
the original large pixel, then the images can be simply interleaved into
a super-image that has small pixels equal to the dither step-size.
If the step-size is small enough, the super-image will be critically sampled.
A simple way to view this is to consider an image consisting of the
astronomical field just convolved with the PSF due to the optics alone.
The sampling would be done on pixels equal to the size of the dither step,
chosen to be fine enough to ensure critical sampling.
The image is then blurred by the original pixel response.
Drawing every $n^{th}$ pixel in $x$ and $y$ clearly recreates
one of the dithered images as actually created by the camera.
Therefore, conversely interleaving the dithered images creates the well-sampled
super-image.

In practice, however, it may not be possible to step the camera in a
regular pattern.  Sub-pixel dithers have
been used in many WFPC-2 programs, for example, but were often
not executed with enough precision to fall on a regular pattern;
simple interlacing of the image-set cannot be done in such cases.
This problem is critical for the {\it Hubble
Deep Field} (HDF) observations \markcite{hdf}(Williams et al.\ 1996).
A regular dither was specified, but did not actually occur.

To solve the problem of combining images with
an irregular dither pattern, a {\it Drizzle}-algorithm was developed
(\markcite{hdf}Williams et al. 1996; \markcite{driz}Fruchter \& Hook 1998)
that works by simply dropping or ``drizzling'' the pixels in any
single image onto a finer grid, offsetting the image by the actual
sub-pixel step obtained, slicing up its pixels as they fall on the finer grid.
The {\it Drizzle} algorithm worked well, producing the now famous well-sampled
full-color image of the HDF.
The {\it Drizzle} algorithm is appealing, as it is intuitive ---
one is just shifting and overlapping the images on a fine grid,
shrinking the original pixels small enough so as to minimize any blurring
associated with forcing the pixels into the new grid, but keeping them
big enough so that there are no ``holes'' of empty data in the new super image.
Further, because {\it Drizzle} works in the spatial domain, it's easy
to correct for cosmic ray events, hot pixels, or any other data missing
in any single images, as well as correcting for any geometric distortion.
Development of {\it Drizzle} represents a significant improvement in
the software available to astronomers for analyzing undersampled images,
and has greatly improved the recovery of information from {\it HST} images.

Despite the success of {\it Drizzle,} however, it
is frankly justified on intuitive rather than formal theoretical grounds,
and indeed depends on two {\it ad hoc} parameters, namely the spacing of the
super-image grid and the size of the pixels to be drizzled.
It also introduces its own blurring function, $\Pi',$ which {\it statistically}
is about the size of the super-image pixel; in detail, the actual resolution
for any object depends on how it falls with respect to the final grid.
Although $\Pi'$ in practice may be much smaller than $\Pi,$ it still may be
large compared to the PSF and introduce significant blurring in its own right.
These issues were indeed discussed in the context of the HDF, and
limit its deconvolution or interpretation of its power spectrum on the
finest scales.

In attempt to develop an algorithm that both mines better
resolution from the data, and stands on a solid theoretical foundation,
I present a method that reconstructs a super-image from an
arbitrary set of dithered observations with no-degradation of resolution.
This method is only a modest extension to two-dimensional
data of a method for recovering one-dimensional functions from
undersampled data presented by \markcite{fft}Bracewell (1978).
The method works by computing the Fourier transform of
the super-image as a linear combination of the transforms of the
individual images; the aliased components are eliminated algebraically.
I have also extended the method to estimate the
super-image when it is actually overdetermined
by the dithered observations.
None of this is particularly complex, and not surprizingly,
the professional image processing literature already
contains discussions of this method (see \markcite{tsai}Tsai
\& Huang 1984, or \markcite{kim}Kim et al. 1990).
However, given the strong interest in using dithers in the context
of {\it HST} imaging, I considered it worthwhile to present this paper
as a tutorial on the method of Fourier algebraic reconstruction
and explore its use in the context of {\it HST} observations.

\section{The Theory of Reconstructing an Image From Aliased Data-sets}

\subsection{The Sampling of a 1-D Function}

\begin{figure}[thbp]
\plotone{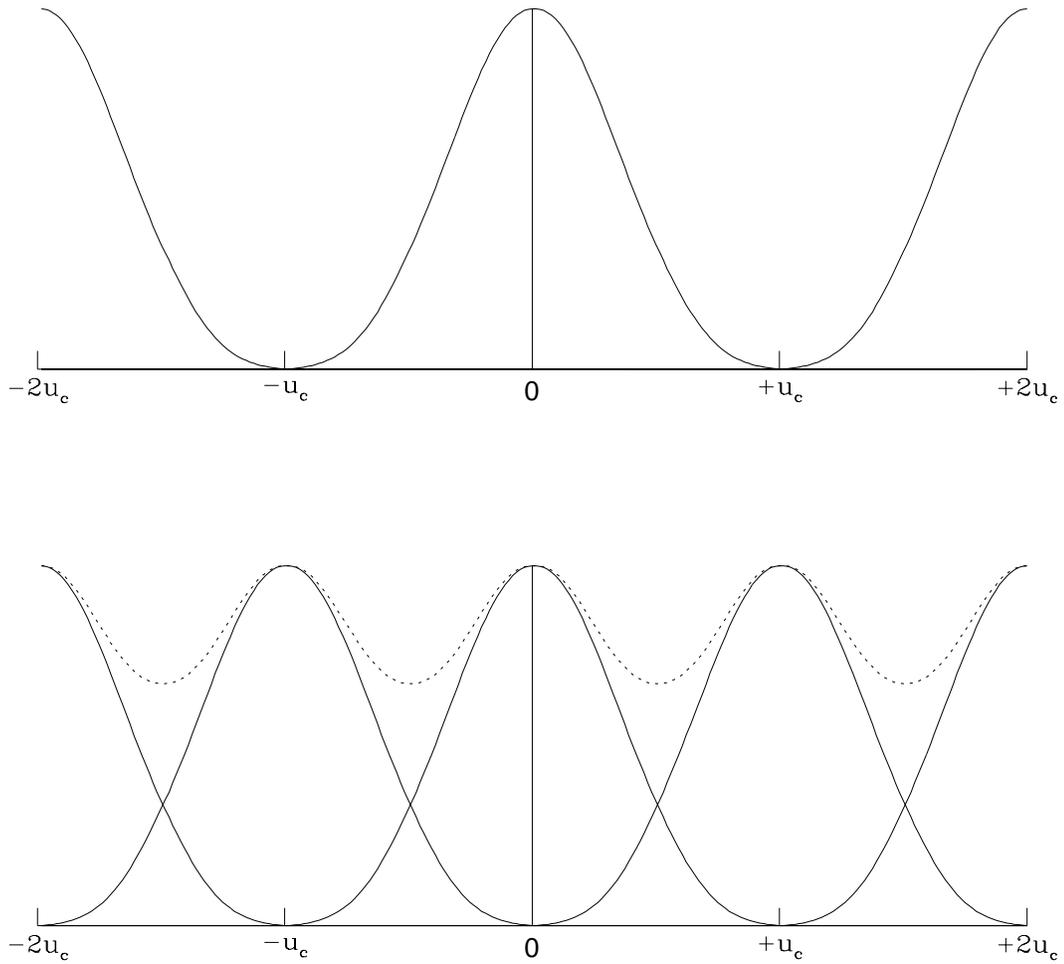}
\caption{This figure schematically shows the effects of sampling
on the Fourier power spectrum of a continuous 1-D function.
Sampling causes the power spectrum to be periodic, with the period
inversely proportional to the spatial sampling frequency.
When the function is well-sampled, the satellites occur at intervals
of $2u_c,$ or greater, where $u_c$ is the critical frequency, or
the highest frequency at which the intrinsic function has non-zero
power (upper graph).  With coarser sampling, the function becomes
undersampled and the satellites begin to overlap.  With $2\times$
undersampling (bottom graph) the satellites occur at every integer
multiple of $u_c.$  The total transform (dotted) is the sum over
all satellites and is severely aliased.}
\label{fig:alias}
\end{figure}

To understand how to reconstruct an image from undersampled data, I start
by considering the effects of sampling on a 1-D function, $f(x).$
For reconstruction to work, $f(x)$ must be band-limited, so that its
Fourier transform,
\begin{equation}
\overline{f(x)}=F(u)=\int_{-\infty}^\infty f(x)e^{-2\pi ixu}dx,
\end{equation}
is non-zero only for $-u_c<u<u_c,$ where $u_c$ is the critical frequency.
If $x$ is expressed in terms of pixels, then sampling at every integer
pixel is sufficient provided that $u_c<1/2.$
This can be understood by
considering the Fourier transform of the sampled function,
The sampling of $f(x)$ is equivalent to multiplying it by a
{\it shah}-function,
\begin{equation}
\shah(ax)\equiv{1\over{\vert a\vert}}\sum_{n=-\infty}^{+\infty}
\delta\left(x-{n\over a}\right),
\end{equation}
where $a=1$ for the specific case of integer-pixel sampling.
The Fourier transform of the sampled function is then,
\begin{eqnarray}
\label{1Dsum}
\overline{f(x)\cdot\shah(x)}&=&F(u)\ast\shah(u),\nonumber \\
&=&{\displaystyle \sum_{n=-\infty}^{+\infty}F(u-n),}
\end{eqnarray}
where I have used the fact that
the transform of a shah-function is itself a shah-function.
As is well-known, the Fourier transform of a sampled function
is periodic, repeating over the entire frequency domain.
If $f(x)$ is band-limited, however, none of the copies or
{satellites} of $F(u)$ overlap.
The satellites are spaced at each integer-step in $u,$
but the requirement that $u_c<1/2,$ means that they also reach
zero before crossing over the midpoint of the interval (Figure \ref{fig:alias}).
This condition is no longer obeyed when the sampling interval is
larger than each integer pixel step.
For example, if every other pixel is sampled, then,
\begin{eqnarray}
\overline{f(x)\cdot\shah\left({x\over2}\right)}&
=&F(u)\ast\shah(2u),\nonumber \\
&=&{\displaystyle 2\sum_{n=-\infty}^{+\infty}F\left(u-{n\over2}\right).}
\end{eqnarray}
The transformed shah-function now samples at every
half-integer step in the Fourier domain, causing strong
overlaps or {\it aliasing} between the satellites of $F(u)$
(Figure \ref{fig:alias}).
If $f(x)$ is unknown, the full extent of its transform cannot be
deduced from the aliased sample, which in turn means that the
sample is itself an incomplete representation of $f(x).$

\subsection{Recovery of a 1-D Function}

\markcite{fft}Bracewell (1978) shows that
a function can be recovered from collection of undersampled data-sets
given prior knowledge of $u_c$ (as might exist given a
detector pixel shape and optical point-spread function), provided
that the sampling among the various data-sets is interlaced by
some fraction of the sampling interval and that the basic sampling
interval is not too sparse compared to $u_c.$
Consider again the alternate pixel sample above (which I relabel
as $d_0(x)$).  For the fundamental interval $-1/2<u<1/2,$
\begin{eqnarray}
D_0(u)&=&{\displaystyle
\overline{f(x)\cdot\shah\left({x\over2}\right)},}\nonumber \\
&=&{\displaystyle {1\over 2}\left(F(u-{1\over2})+
F(u)+F(u+{1\over2})\right).}
\end{eqnarray}
Since I have specified that $F(u)$ is band-limited to $\vert u\vert<1/2,$
for $0\leq u<1/2,$
\begin{equation}
D_0(u)={1\over 2}\left(F(u-{1\over2})+F(u)\right),
\end{equation}
and for  $-1/2<u<0,$
\begin{equation}
D_0(u)={1\over 2}\left(F(u)+F(u+{1\over2})\right).
\end{equation}
Now let there be a second data set that also samples $f(x)$ with
alternate pixel spacing, but spatially offset from the $d_0(x)$ samples
by some $x_0\neq2n$ (one might presume
$0<x_0<2,$ but this is not required).
The transform of the new data-set, $d_{x_0}(x),$ is
\begin{eqnarray}
D_{x_0}(u)&=&{\displaystyle
\overline{f(x)\cdot\shah\left({x\over2}-x_0\right)},}\nonumber \\
&=&{\displaystyle
F(u)\ast\overline{\left(\shah
\left({x\over2}\right)\ast\delta(x-x_0)\right)},}\nonumber \\
&=&{\displaystyle
F(u)\ast\left(\shah\left(2u\right)\cdot e^{-2\pi i ux_0}\right).}
\end{eqnarray}
This reduces to
\begin{eqnarray}
D_{x_0}(u)&=&{1\over2}\left(F(u)+e^{-\pi i x_0}
F(u-{1\over2})\right),\qquad 0\leq u<1/2, \\
&=&{1\over2}\left(F(u)+e^{+\pi i x_0}F(u+{1\over2})\right),
\qquad -1/2<u<0.
\end{eqnarray}
Note that $d_{x_0}(x)$ is no less aliased than is $d_0(x),$
but since the overlap
portion has a differing phase, the transforms of the two samples
can be combined to solve for the transform of $f(x),$
\begin{eqnarray}
F(u)&=&{\displaystyle 2~{D_{x_0}(u)
-e^{-\pi i x_0}D_0(u)\over
1-e^{-\pi i x_0}},}\qquad 0\leq u<1/2, \\
&=&{\displaystyle 2~{D_{x_0}(u)-e^{+\pi i x_0}D_0(u)\over
1-e^{+\pi i x_0}},}\qquad -1/2< u<0.
\end{eqnarray}
In other words, one can reconstruct $f(x)$ exactly from two data-sets
offset from each other, each undersampled by a factor two.
Note that in the special case, where $x_0=1,$ $d_0(x)$
holds the even-numbered pixels and $d_{x_0}(x)$ holds the odd-numbered ones,
then
\begin{equation}
F(u)=D_{x_0}(u)+D_0(u),
\end{equation}
as would be expected, since the sum in equation (\ref{1Dsum}) can clearly
be separated this way.  With exact interlacing, one can just add the
transforms of the two individual data-sets (provided that the transform
preserves their relative phases).

\subsection{Recovery of an Image}

\begin{figure}[thbp]
\plotfiddle{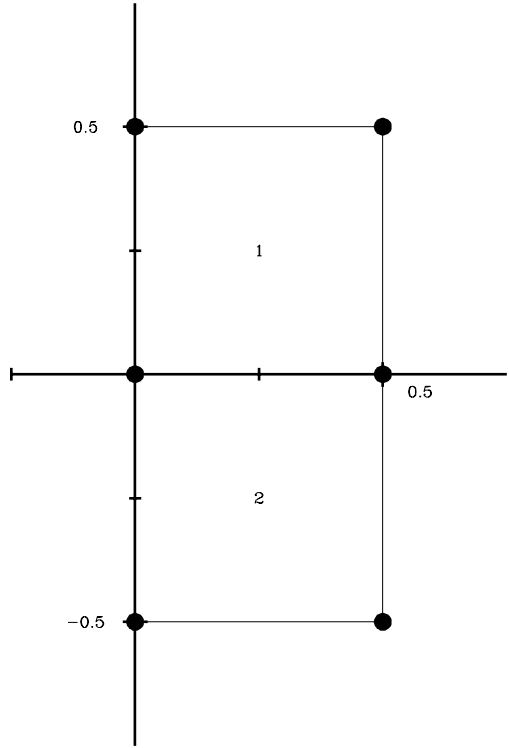}{2.5in}{0}{100}{100}{-180}{-144}
\caption{This figure schematically shows the configuration of the Fourier
domain for reconstructing an image with $2\times2$ subsampling.
For real images, Fourier transforms need only be calculated
for the semi-plane with $0\leq u\leq1/2,$ $-1/2<v\leq1/2$
(this presumes that the $x$-axis transform is computed first),
where the frequencies are defined with respect to the pixels of the
reconstructed image.  Each image in the observed set is aliased, and
has satellites at all integer multiples of $(u,v)=1/2$ in the Fourier
domain, with each satellite having significant power over $\Delta u=\pm1/2,$
and $\Delta v=\pm1/2$ about its central location.  The figure
shows as heavy dots the central location of all satellites that overlap
with the fundamental transform centered at $(u,v)=0.$
Algebraic elimination of the satellites is done in two regions, marked
1 and 2; the satellites that contribute to a given region are those
at its corners.}
\label{fig:2x2}
\end{figure}

This method can be directly generalized to the case of reconstructing
a 2-D image.  The shah-function becomes a 2-D regular grid of
$\delta$-functions, and the two-dimensional Fourier transform 
of an image is:
\begin{equation}
\overline{f(x,y)}=F(u,v)=\int_{-\infty}^\infty
\int_{-\infty}^\infty f(x,y)e^{-(2\pi ixu + 2\pi iyv)}dx~dy.
\end{equation}
If there is an observation $d_{{x_1},{y_1}}(x,y)$ that is factor of
two undersampled in both $x$ and $y$ (thus having 1/4 of the pixels
of the critically sampled image), and offset by $x_1,\ y_1$ from the
nominal grid defining $f(x,y),$
then in the domain $0\leq u<1/2,\ 0\leq v<1/2,$
\begin{eqnarray}
D_{{x_1},{y_1}}(u,v)&=&{1\over4}\bigg(F(u,v)+e^{-\pi i x_1}
F(u-{1\over2},v) \nonumber \\
&&+e^{-\pi i y_1}F(u,v-{1\over2})+e^{-\pi i (x_1 + y_1)}
F(u-{1\over2},v-{1\over2})\bigg).
\end{eqnarray}
There are analogous expressions in the other three
quadrants of the $u,v$ plane;
however, for real-valued images, half of the $u,v$ plane will simply
be the complex conjugate of the other half and thus need not be computed
(see Figure \ref{fig:2x2}).
As can be seen, with four data-sets, each having a unique offset in
$x$ {\it or} $y,$ it is again possible to eliminate the overlap
contributions.  This requires solving a system of equations with
complex coefficients:
\begin{equation}
{1\over4}\left(\matrix{1&e^{-\pi i x_1}&e^{-\pi i y_1}&e^{-\pi i(x_1+y_1)}\cr
1&e^{-\pi i x_2}&e^{-\pi i y_2}&e^{-\pi i(x_2+y_2)}\cr
1&e^{-\pi i x_3}&e^{-\pi i y_3}&e^{-\pi i(x_3+y_3)}\cr
1&e^{-\pi i x_4}&e^{-\pi i y_4}&e^{-\pi i(x_4+y_4)}\cr}\right){\bf F}={\bf D},
\end{equation}
where ${\bf F}$ is a 4-vector holding $F(u,v)$ in the first position,
followed by the $u,$ $v,$ and lastly $u,v$ satellites, ${\bf D}$
is a 4-vector of the transforms of the 4 undersampled data-sets.
One can then invert this matrix to find
\begin{equation}
\label{comb}
F(u,v)=\sum_{n=1}^4c_nD_{{x_n},{y_n}}(u,v),
\end{equation}
where $c_n$ will be a complex coefficient.
Solution for the second quadrant is analogous --- the phases
differ only in sign, being positive when the domain of the frequency
is negative.
As an example, for the special case of where the four data-sets
contain the exact interlaces of integer pixels in $x$ and $y,$
{\bf F} and {\bf D} are more simply related as:
\begin{equation}
{1\over4}\left(\matrix{\phantom{-}1&\phantom{-}1&\phantom{-}1&\phantom{-}1\cr
\phantom{-}1&-1&\phantom{-}1&-1\cr
\phantom{-}1&\phantom{-}1&-1&-1\cr
\phantom{-}1&-1&-1&\phantom{-}1\cr}\right) {\bf F}={\bf D},
\end{equation}
which has the solution, as expected of
\begin{equation}
F(u,v)=\sum_{n=1}^4D_{{x_n},{y_n}}(u,v).
\end{equation}

\subsection{Recovery of an Image Overdetermined by the Data}

Four images determine $F(u,v),$ exactly, but
if one actually has additional images available,
$F(u,v)$ is overdetermined, and a least squares solution is required.
This means finding the $F(u,v)$ that minimizes the norm
\begin{equation}
E=\Vert{\bf\Phi}{\bf F}-{\bf D}\,\Vert,
\label{norm}
\end{equation}
where, as above ${\bf\Phi}$ is the matrix of phases.
In this case, however, ${\bf\Phi}$ is now an $n\times4$ matrix,
\begin{equation}
{\bf\Phi}={1\over4}\left(
\matrix{1&e^{-\pi i x_1}&e^{-\pi i y_1}&e^{-\pi i(x_1+y_1)}\cr
1&e^{-\pi i x_2}&e^{-\pi i y_2}&e^{-\pi i(x_2+y_2)}\cr
\vdots&\vdots&\vdots&\vdots\cr
1&e^{-\pi i x_n}&e^{-\pi i y_n}&e^{-\pi i(x_n+y_n)}\cr}\right),
\end{equation}
where $n\geq4$ is the number of data-sets, and ${\bf D}$ is now a vector of
length $n$ holding the data-sets; ${\bf F}$ is still the same 4-vector.
Expanding equation (\ref{norm}) gives
\begin{eqnarray}
E^2&=&\left({\bf\Phi}{\bf F}-{\bf D}\,\right)^H
\left({\bf\Phi}{\bf F}-{\bf D}\,\right)\nonumber \\
&=&{\bf F}^H{\bf\Phi}^H{\bf\Phi}{\bf F}-{\bf F}^H
{\bf\Phi}^H{\bf D}-{\bf D}^H{\bf\Phi}{\bf F}+{\bf D}^H{\bf D},
\end{eqnarray}
where $H$ denotes the Hermitian (or complex-conjugate) transpose.
Minimizing $E$ implies
\begin{equation}
\label{F2D}
{\bf F}=\left({\bf\Phi}^H{\bf\Phi}\right)^{-1}{\bf\Phi}^H{\bf D}.
\end{equation}\par
In the case of an overdetermined situation, one might further
want to weight the observations differently.
For example, it may not be practical to obtain exposures of identical
length over the sequence of observations, or they may have variable
backgrounds.  In this case, it's easy to generalize equation (\ref{F2D})
to include weighting, giving
\begin{equation}
{\bf F}=\left({\bf\Phi}^H{\bf W}^T{\bf W}{\bf\Phi}\right)^{-1}
{\bf\Phi}^H{\bf W}^T{\bf W}{\bf D},
\label{F2W}
\end{equation}
where ${\bf W}$ is an $n\times n$ matrix of weights and ${\bf W}^T$
is its transpose (the weights are real-valued).
${\bf W}$ can account for any covarience between the images,
but it is most likely to be diagonal on the presumption that the
individual images will probably be independent.

\subsection{Generalization to Higher Degress of Subsampling}

\begin{figure}[htbp]
\plotfiddle{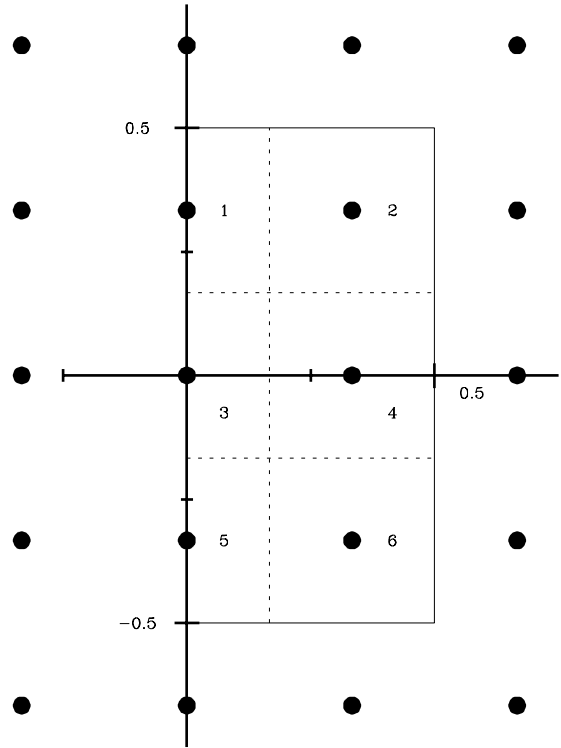}{3.0in}{0}{100}{100}{-162}{-144}
\caption{As for the $2\times2$ case,
this figure schematically shows the configuration of the Fourier
domain for reconstructing an image with $3\times3$ subsampling.
Again, the Fourier transforms are calculated
only for the semi-plane with $0\leq u\leq1/2,$ $-1/2<v\leq1/2.$
Satellites now occur at all integer multiples of $(u,v)=1/3,$
but each satellite still has significant power over $\Delta u=\pm1/2,$
and $\Delta v=\pm1/2$ about its central location.  The figure
shows as heavy dots the central location of all satellites that overlap
with the fundamental transform centered at $(u,v)=0.$
Algebraic elimination of the satellites is now done in six regions;
the satellites that contribute to a given region are the one at its center,
and the eight surrounding it.}
\label{fig:3x3}
\end{figure}

Double sampling is likely to be sufficient to remove modest aliasing,
but higher levels of subsampling may be required
when the undersampling is severe.
Generalization to finer levels of subsampling is straight forward,
if somewhat tedious.  As the observed images become coarser with respect
to the reconstructed image, the aliased satellites become closer
together and overlap more severely.  Algebraic elimination
of the satellites requires identifying
all satellites contributing power to a given location in the
Fourier domain.  In practice, this means slicing the Fourier
domain into an increasingly large number of subsets.
Figure \ref{fig:3x3} sketches out the structure of the Fourier
domain for $3\times3$ subsampling.
In the $3\times3$ case, the Fourier domain is divided into six regions,
with nine differing satellites contributing to ${\bf F}$ in each one;
at least $n\geq9$ dithered
images will be required to find a solution, and 
${\bf\Phi}$ is will now be an $n\times9$ matrix.
An important distinction between the $2\times2$ and $3\times3$ cases,
is that in the former, since the satellites are spaced exactly by
$u_c,$ only the six satellites that are visible within the
Fourier semi-domain need be considered.  In the $3\times3$ case,
the satellites are separated only by multiples of $2u_c/3,$ thus
the first set of satellites with their {\it centers} actually falling
outside the semi-domain will still overlap with it.

\section{Implementation of the Fourier Image Reconstruction}

\subsection{Data-set Requirements\label{sec:req}}

The present reconstruction method works
only if the data satisfies a number of conditions,
the most important of which is that the intrinsic image
structure remain constant over the extent of the dithered data-taking sequence.
The PSF should not vary significantly in time, or if the
dither steps are large, in space as well.
``Significantly'' in this context means variations on spatial scales
where the Fourier $S/N$ ratio is greater than unity;
bright point sources are more vulnerable to PSF-variations than
faint or more diffuse sources.
Bright noise spikes, hot pixels,
cosmic ray hits, or any other variable sources, must also be eliminated or
repaired prior to reconstruction.
A final obvious requirement is that reconstruction can work only
on the portions of the dither set in common to all images;
as the dither takes place, it is likely that a larger region of
the sky will be imaged than is present on any single image ---
subimages of the common overlap region must be isolated prior to reconstruction.

The mathematics of the Fourier reconstruction method do not strictly require
that the angular size of the pixels be constant over the extent
of any image, provided that the dither steps are small enough that
they can be regarded as constant over the complete area of the images.
Images that have variations in their pixel scale large enough so that
the amplitude of the dithers (in pixels) varies significantly over
the extent of the image must be processed in subsets small enough
that the dithers can be regarded as constant over the angular domain selected.
Lastly, the dithers must be translational only, with no rotation.

The reader familiar with {\it Drizzle} may object
that these requirements are too restrictive for many sets of dithered data.
{\it Drizzle} performs cosmic ray event and defect rejection,
as well as geometric rectification, when building a sub-sampled image.
{\it Drizzle} is thus attractive for the complete reduction of
panoramic data sets.  This issue will be discussed further in $\S\ref{sec:sum},$
but I emphasize that the present approach is solely concerned with
the specific task of accurate reconstruction of a Nyquist-sampled image.
Geometric rectification or defect rejection are problems
that can be separated from the actual reconstruction algorithm;
the caveats presented above do not necessarily prevent use of the
present method if they can be addressed apart from the reconstruction task.

Two other requirements on the data set concern the pattern and measurement
of the dithers.  Ideally, the fractional portion of the dither steps (that
is ignoring the integer number of pixels stepped over) should match
the nominal $2\times2$ or $3\times3$ equal sub-stepping patterns as
closely as possible; or if the problem is heavily overdetermined be
at least evenly spread over the area of a single pixel.
In this case, solution of equation (\ref{F2W}) will generate a set of complex
coefficients, $c_n$ of nearly equal power (presuming
equal weights).
Formally solutions can be calculated for any nondegenerate dither pattern;
however, as the dither pattern moves away from optimal, the
images will be combined unevenly, with heavy weight being placed on
those with less redundant positions.
For real images, this means that the relative noise contributed
by such images will be amplified compared to others in the dither set.
Noise properties of the reconstructed image will be discussed below;
in practice, excess amplification of noise is only important for large
departures from an ideal pattern.

Accurate measurement of the dither steps is required to construct
the $\Phi$ matrix.  This may be done iteratively.
Initially one might use simple centroids of stars or other compact
objects within a given image to measure dither offsets.
Once a reconstructed image has been generated,
it can be cross-correlated with the individual images to refine the offsets;
permitting a more accurate reconstruction to be done in a second iteration.

\subsection{Computing the Reconstructed Image}

Given the prepared set of dithered images and measured dither steps,
computation of the reconstructed image can proceed.
In practice I have done this within the Vista image processing system,
making use of its native image arithmetic and Fourier routines,
augmenting it only with a new subroutine to construct $\Phi,$ and
then solve for and apply $c_n$ to the Fourier transform of a given image.

For each image, the first steps are to normalize it
to a common exposure level, and to then expand it into a sparse array,
spacing out the pixels by $2\times$ or $3\times$ as desired.
Each pixel in an input image then occupies one of the corners of 
a cell of $2\times2$ or $3\times3$ new pixels in the expanded image,
with the other $n\times n-1$ pixels in each cell set to zero.
This actualizes each image as a sparse \shah\ function;
one can see that for exact $n\times n$ dithers,
the other images would simply be interlaced at the vacant locations.

Once an image is expanded, its Fourier transform is computed;
a power spectrum at this stage clearly shows the aliased satellites.
The next step is to multiply the transform by $c_n$, remembering that
different coefficients must be used for the various regions
within the domain.  The adjusted transform is then added to
the adjusted transforms of the other images.
The reconstructed image is the inverse transform of the complete sum.

One important caveat is that each the transform of each image must be
multiplied by a complex phase, $\exp\left(-2\pi Ki\left(x_j+y_j\right)\right),$
where $(x_j,y_j)$ is its spatial offset from the average of the other images,
and $K$ is the degree of subsampling.  This is required
because the mathematics presented in the previous section presume
a two-dimensional coordinate system anchored to the sky, rather
than the grid of the detector.
In other words, as each image is expanded, initially its \shah\ function
has identical coordinates to those in the other images, with the object
apparently moving with respect to the detector coordinate system.
This step resets the coordinate system to that of the sky,
correctly phasing the various \shah\ functions of the dither set.

\subsection{Examples of Reconstructed Images}

\begin{figure}[htbp]
\plotone{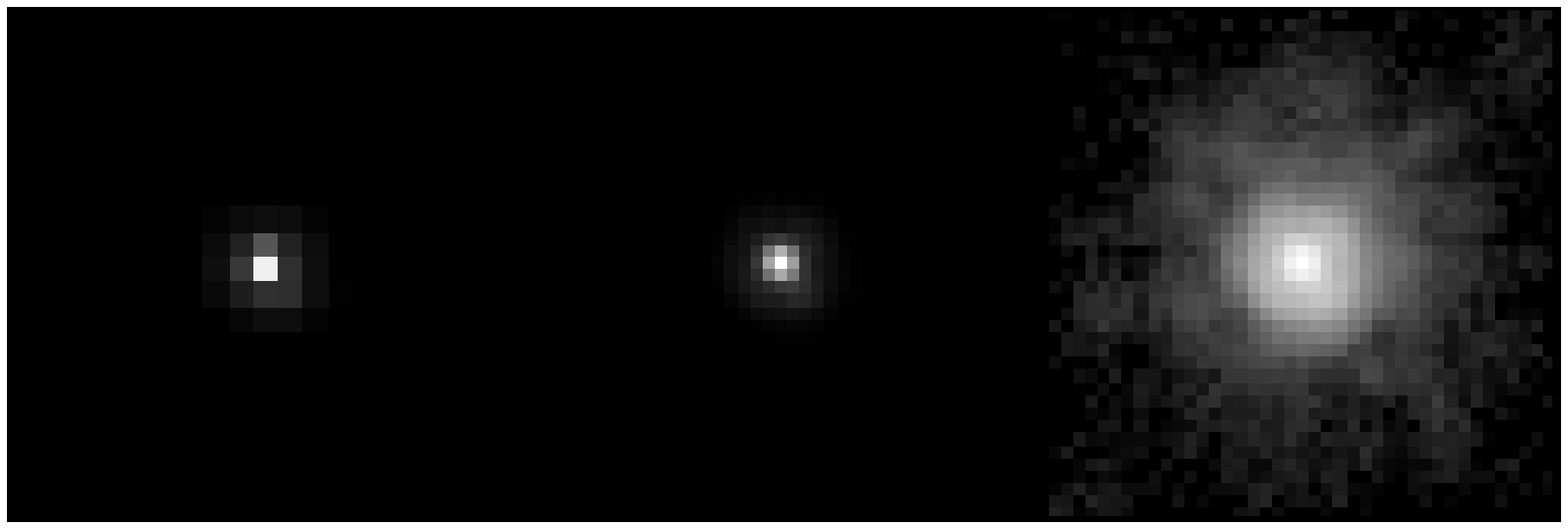}
\caption{Reconstruction of the {\it HST} PC PSF with $2\times2$
subsampling is shown based on 20 dithered F555W images of a
star in $\omega$ Cen.  The image at left shows a linear stretch of one
of the PSF images (selected to be nearly centered on a pixel).
The central image shows the reconstructed PSF with the same intensity stretch.
The last image is a logarithmic stretch (with dynamic range 3.5 in log units)
of the reconstructed PSF.}
\label{fig:pc_psf}
\end{figure}

\begin{figure}[hbtp]
\plotone{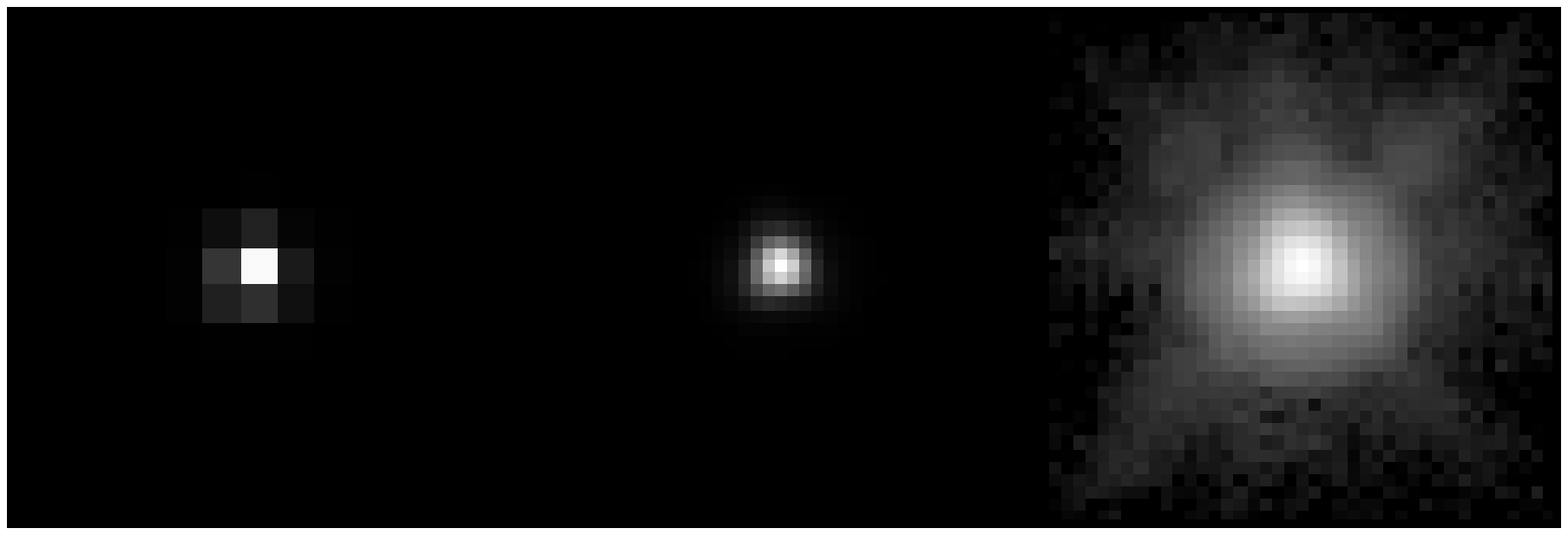}
\caption{Reconstruction of the {\it HST} WFC PSF with $3\times3$
subsampling is shown based on 20 dithered F555W images of a
star in $\omega$ Cen.  The image at left shows a linear stretch of one
of the PSF images (selected to be nearly centered on a pixel).
The central image shows the reconstructed PSF with the same intensity stretch.
The last image is a logarithmic stretch (with dynamic range 3.5 in log units)
of the reconstructed PSF.}
\label{fig:wfc_psf}
\end{figure}

Figures \ref{fig:pc_psf} and \ref{fig:wfc_psf} show PC and WFC PSFs
reconstructed from a calibration program of 20 F555W dithered images of
a field within the $\omega$ Cen globular cluster.
The PC PSF was reconstructed with $2\times2$ subsampling, while
$3\times3$ subsampling was used for the WFC PSF.
The cores of the PSFs are now well resolved, and no ``boxy'' artifacts
are seen as can occur in {\it Drizzle} reconstructions
(\markcite{driz}Fruchter \& Hook 1998).
It's also worthwhile to note the strong blurring introduced by the
WFC pixel function, $\Pi,$ itself.
Again, the reconstruction does not recover the intrinsic PSF due to
the optics only, but the intrinsic PSF convolved with $\Pi.$
The PC PSF clearly has the sharper and rounder core, while the
center of the WFC PSF is strongly determined by the pixel shape.

\begin{figure}[thbp]
\plotone{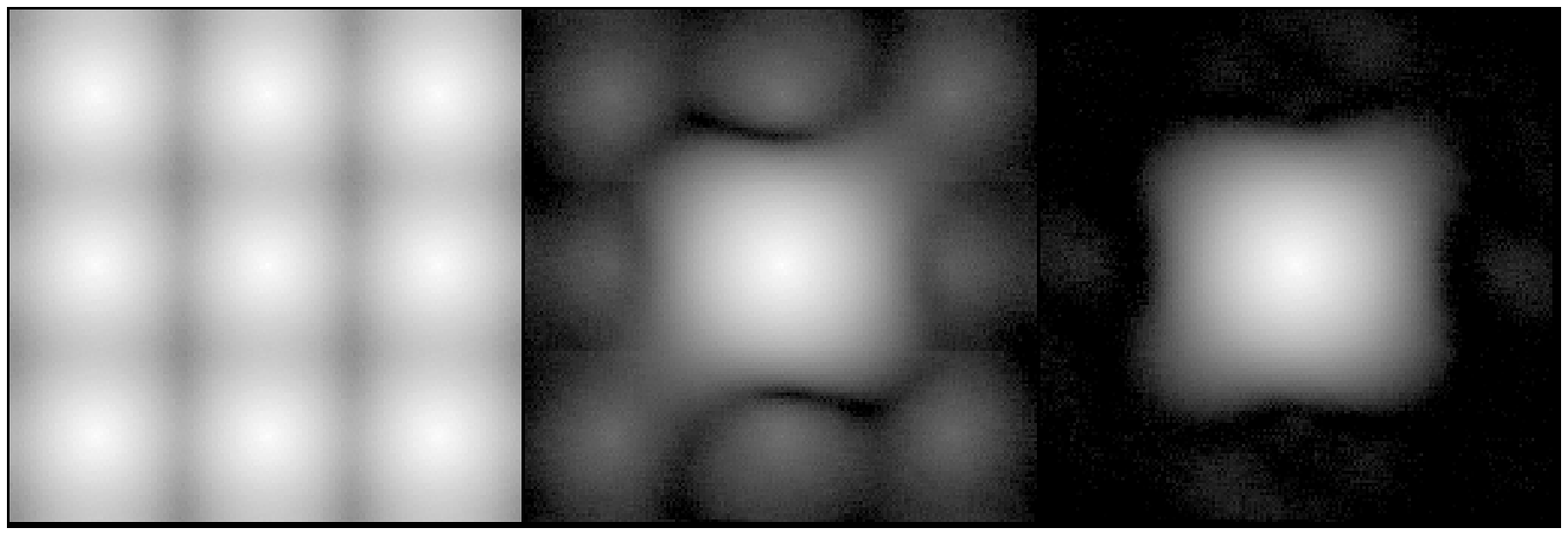}
\caption{Power spectra are shown at various stages in the reconstruction
of the WFC PSF with $3\times3$ subsampling.  The left image shows the
power spectrum of a single PSF image expanded as a sparse $\shah\ $ function.
The low contrast of the minima between the bright peaks of the satellites
shows the effects of the severe aliasing in WFC images.  The middle
image shows the spectrum of the penultimate reconstruction.  At this
stage 19 of the 20 images have been combined and the flanking satellites
have been greatly reduced in power.  The right image shows the power spectrum
of the final reconstructed PSF --- the partial combination shown in the
middles has now been completed by the addition of the last image. The
display scale is identical and logarithmic (with a range of $10^5$) for
all three spectra. The power spectra are shown for the full Fourier
domain for ease of visual interpretation, even though the transforms
are computed only in a semi-plane.}
\label{fig:power}
\end{figure}

Figure \ref{fig:power} shows the power spectra at various stages in the
reconstruction of the WFC PSF to illustrate
\begin{figure}[thbp]
\plotone{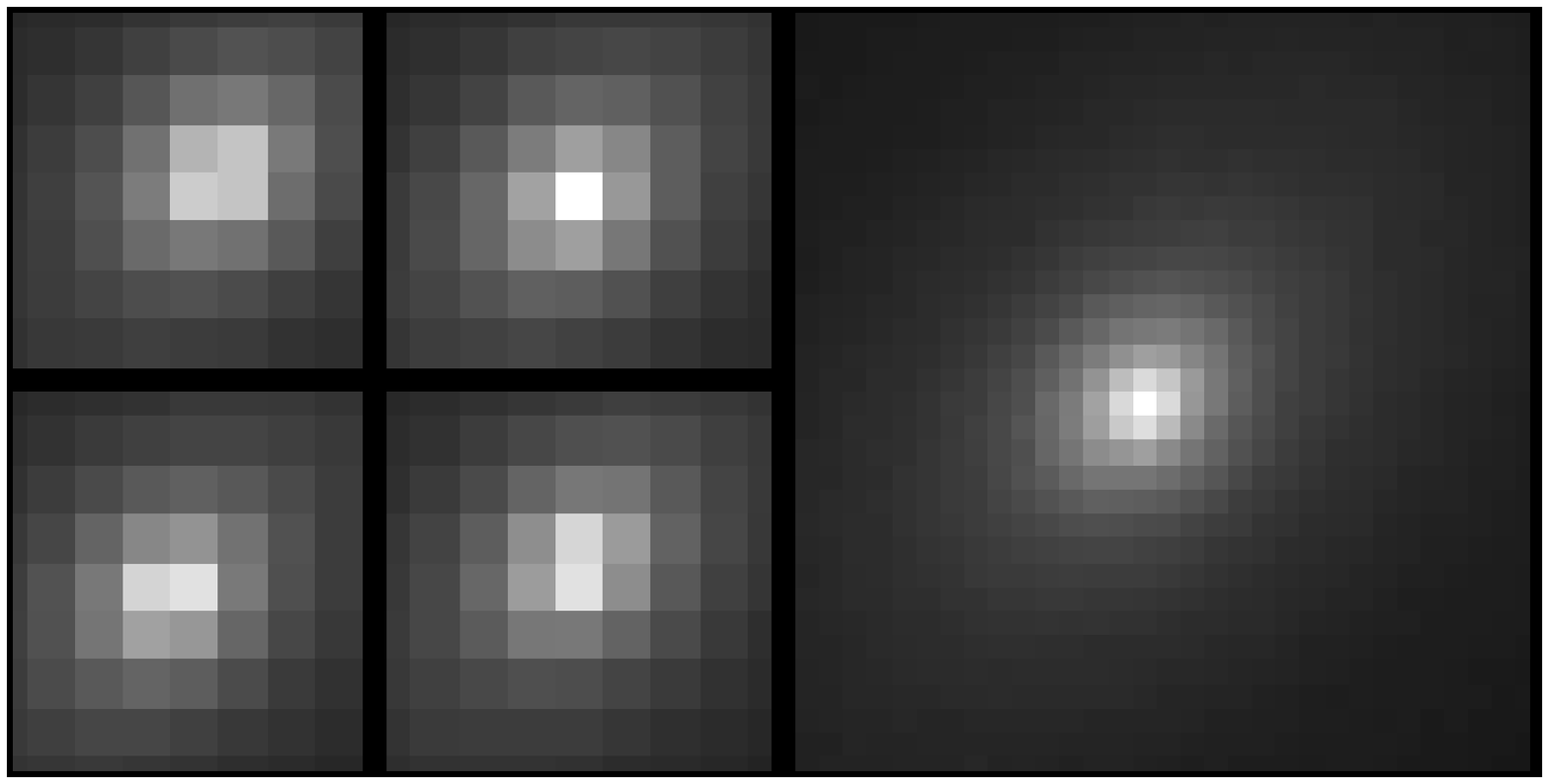}
\caption{The reconstruction of the center of NGC 1023 with $2\times2$
subsampling.  Five F555W PC images were used.  Four of the images
(shown at left) define an approximate $2\times2$ interlace pattern;
however, the offsets typically differed from the nominal 0.5 pixel
steps by $\sim0.1$ pixel (the fifth image falls within 0.1 pixel of one
of the four images shown).  The stretch is linear.}
\label{fig:n1023}
\end{figure}
the algorithm concretely.  The final combination of 20 images has
reduced the contribution of the aliased satellites by $\sim10^5.$
The final power spectrum also ratifies the strong contribution
of the WFC pixel to the total PSF.  The shape of the spectrum is
clearly boxy; further, the central lobe is surrounded by a strong
zero, which would be expected in the power spectrum of a nearly
square and uniform pixel function.

Turning to more interesting objects,
Figure \ref{fig:n1023} shows the $2\times2$ reconstruction of the
nucleus of the early-type galaxy NGC 1023.
Unlike the situation for the PSFs, which were highly overdetermined,
only five dithered images were available for NGC 1023.
The dither pattern was close to a nominal exact interlace,
but the offsets typically differed from the nominal 0.5 pixel
by $\sim0.1$ pixel, thus the present method was required.
This galaxy has a particularly compact center (\markcite{l95}Lauer et al. 1995).
The present observations were obtained to observe its central structure
with the best resolution available --- reconstructing the image without
introducing additional blurring is thus critical.
The reconstructed image clearly shows the sharp compact nucleus of NGC 1023,
but is also smooth and free from artifact; indeed this image can now be
processed further with PSF deconvolution.

\begin{figure}[thbp]
\plotone{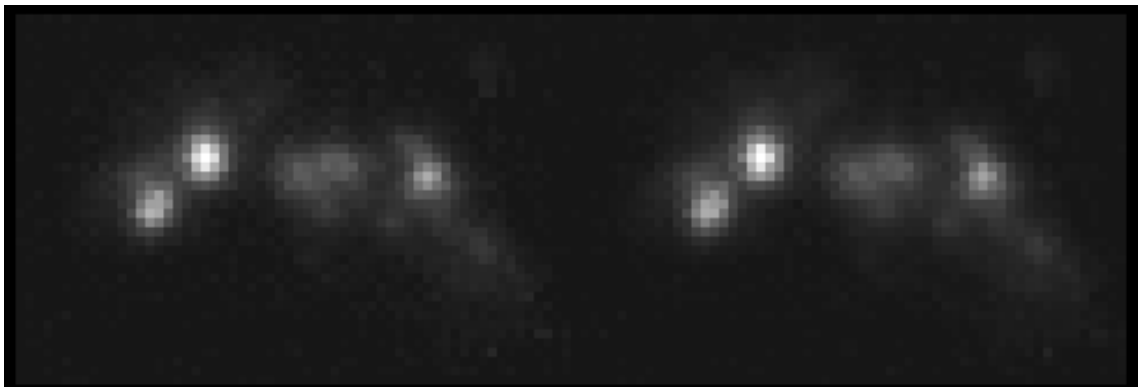}
\caption{Two reconstructions of a $z=1.355$ chain galaxy in
the {\it Hubble Deep Field} with $2\times2$
subsampling, based on 11 F450W WFC images.  The left image
was done with the present Fourier method, while the image on the
right is a {\it Drizzle} reconstruction.  The stretch is linear.}
\label{fig:hdf_gal}
\end{figure}

Lastly, I show a $2\times2$ reconstruction of a chain-galaxy at
$z=1.355$ (\markcite{cohen}Cohen et al. 1996) in the {\it Hubble Deep
Field} (Figure \ref{fig:hdf_gal}), along with a {\it Drizzle}
reconstruction.
\footnote{The {\it Drizzle} reconstruction shown was done with the
same image set, weights, and pixel grid used for the Fourier reconstruction,
and differs from the {\it Drizzle}-reconstructed image of the
same galaxy in the official release of the {\it HDF.}}
Superficially the two images look identical; the gross morphology
is not strongly dependent on the reconstruction algorithm.
Detailed comparison shows, however, that the present reconstruction
is slightly sharper --- the peak of the brightest knot in the image
is $\sim7\%$ brighter, for example.
Matching the resolution of the {\it drizzled} image requires
smoothing the Fourier reconstruction with a Gaussian with $FWHM\approx1$
pixel (on the subsampled scale).  The Fourier reconstruction
does appear to have more noise, but again this is due to the
smoothing inherent in the {\it Drizzle} algorithm.
The Fourier reconstruction can be smoothed, but
one of the nice things about having a well-sampled image is that
optimal filters can be used to improve its appearance.
A Weiner filter, for example, can be used to reject much of the
noise in the present image with little effect on its resolution;
an option that is not possible with aliased images.

A more general comparison of the present method to {\it Drizzle} is
complex, as the difference between the two depends on the dither
pattern, the size of the image set, choice of the
reconstructed pixel size, and the {\it Drizzle} drop size.
For example, when the dither pattern is close to an exact interlace,
{\it Drizzle} can be configured to produce a simple interlaced
reconstruction, while at the opposite end of the scale, {\it Drizzle}
can do simple ``shift-and-add'' reconstructions on the
original pixel scale, which implies highly significant smoothing.
In general, it appears from a number of
additional experiments that when a large image set is available,
{\it Drizzle} effectively smooths a perfect
reconstruction with a gaussian with width of about one pixel, as
in the HDF galaxy above.
For WFC PSFs, for example, the blurring can cause a 10\%\ reduction
in the flux of the central pixel.
This is not guaranteed, however; in one WFC PSF experiment with only four
nearly exactly interlaced images, {\it Drizzle} produced a result
that was apparently {\it sharper} than the Fourier reconstruction.
Close examination, however, showed that the {\it Drizzle} result
was still aliased; aliasing can cause features to be artificially
sharpened as well as broadened.  Further comparison of the
Fourier method to {\it Drizzle} is thus best done in a context
specific to the scientific problem at hand.

\subsection{Noise in the Reconstructed Image}

As alluded to in $\S\ref{sec:req},$ the noise level in
the reconstructed image depends on how well the dither pattern
matches an ideal interlace pattern.  For $N$ images, the solutions
presented in equations (\ref{F2D}) or (\ref{F2W}) reduce to a
set of complex coefficients $\{c_n\}$ relating $F(u,v)$ to the
data, as in the exact solution shown in equation (\ref{comb}).
On the presumption that the noise from image to image is uncorrelated,
then the average power in noise in the reconstructed image is simply
\begin{equation}
\label{noise_lev}
\eta_F=\left(\sum_{n=1}^Nc_n^* c_n\eta_n^2\right)^{1/2},
\end{equation}
where $\eta_n$ is the noise power in image $n.$
With a nearly ideal dither pattern (and equally-weighted
data), $(c_n^*c_n)^{1/2}\approx K^2/N,$
where $K$ is the degree of subsampling; the noise level is as
expected for the simple addition of $N$ images.
As the dither pattern becomes less ideal, however, unequal weight
is placed on the images, depending on the uniqueness of their positions.
Highly redundant images will have small coefficients, while
more isolated images contribute relatively higher power.
The linear combination of the images still produces an exact
solution for the reconstructed image, but because the noise is incoherent from
image to image it may be amplified in the final image, relative
to its level in the ideal case.
Equation (\ref{noise_lev}) allows the noise in the reconstructed
image to be calculated in advance for any particular dither pattern.

\begin{figure}[hbtp]
\plotone{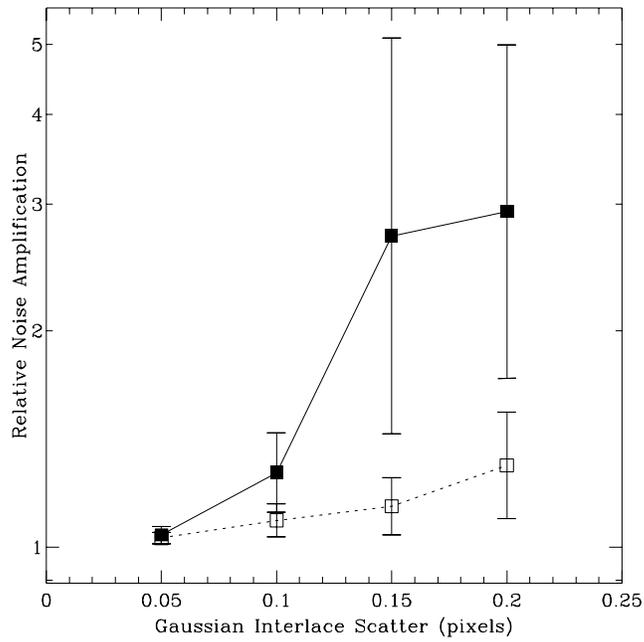}
\caption{Simulations of the relative amplification of noise as a function
of the departure of the dither pattern from a perfect interlace are shown.
The departure is parameterized as a normal distribution of random
offsets with the standard deviation specified in original pixels.
The solid curve and points show the case for when only four images
are used in the reconstruction.  The dashed-line and open symbols
show the simulations done with nine input images.}
\label{fig:noise}
\end{figure}

Figure \ref{fig:noise} gives shows how the noise level in the reconstructed
image varies as the dither pattern moves away from the ideal interlace
for two examples of $2\times2$ subsampling.  In these tests,
variations in the dither pattern were treated as random gaussian errors
about the exact interlace.  For a given standard deviation of the
random offset, several simulated image reconstructions were computed.
For the example with only four images, there is no redundant information,
and the noise level depends strongly on the particulars of the dither
pattern once excursions from the exact interlace become large.
For nine images the reconstruction is more stable to
departures from the ideal pattern, the final noise level showing less
large excursions.  The real importance of this demonstration, however,
is to show that the noise level rises only slowly above its ideal for
small errors in the dither pattern.  Experience with WFPC-2,
for example, shows that typical dithering errors
($\lesssim0.1$ PC pixel)
will give results within the regime of modest noise amplification.

\section{Discussion and Summary\label{sec:sum}}

As noted in the introduction, my interest in the Fourier reconstruction method
presented here stemmed from a strong desire to avoid the random blurring,
$\Pi',$ that {\it Drizzle} may introduce into the reconstructed image.
The present method permits exact reconstruction of the superimage, with no
blurring at the Nyquist scale, nor requires any arbitrary decisions or
parameters to control the form of the reconstructed image.
One might object that the degree
of subsampling selected is such a parameter; however, it is really
specified by the intrinsic spatial scale of the Nyquist frequency.
A Nyquist-sampled image can be resampled at finer scales without loss
of information content or introduction of artifact --- images generated
at various subsampling scales past the Nyquist scale are essentially
equivalent representations of the image.

The present algorithm places several preconditions on the data,
thus it is worthwhile to consider 1) the optimal data-taking strategy for
reconstructing images from dithered data-sets, and 2) how to best perform
the related tasks of artifact rejection, geometric rectification, and so on.
The mathematics of the Fourier method strongly recommends selecting
a dither pattern that contains fractional offsets as close to the
ideal interlace pattern, itself.  If a good dither pattern is realized,
little is demanded of the linear combination of the images --- one
is simply accounting for the slight errors in its execution.  It
should be emphasized that the dither pattern can also contain
integer pixel offsets as well, as might be desired to eliminate
hot pixels, traps, blocked columns, and other fixed detector defects
as well as cosmic rays.  A nearly ideal program for the present
algorithm would be to attempt a $2\times2$ subsampling interlace,
but taking multiple exposures at each dither step to allow for
cosmic ray rejection.  This strategy clearly demands a rather large
data-set, which may not be feasible for programs lasting only
an orbit or two on {\it HST.}  However, it presents no difficulties
for multi-orbit programs, where one will be obtaining a large number
of exposures in any case.

With regards to the second issue above, I have focused solely on
the problem of reconstructing a Nyquist-sampled image.  Tasks that are
required before this stage include image registration and defect repair.
Tasks that might follow reconstruction include geometric rectification,
deconvolution, and filtering.
{\it Drizzle} is attractive in part because it is a complete package that
does many of these steps together within the familiar IRAF/STSDAS environment.
This said, however, I emphasize that many of the preliminary reduction
steps can be done independently of the Fourier reconstruction
algorithm --- these issues should not impede its use.
Indeed, one might use {\it Drizzle} for an initial reconstruction
to provide for defect rejection prior
to a second reconstruction cycle using the present algorithm.
Geometric rectification is simple in principle if one is working with
well-sampled images; the issue is generating such an image if geometric
distortions are important in the undersampled observations.
As noted earlier, if the dithers are small, scale changes across the
image may be unimportant; if variations in the local dither step over
the image domain are limited to a few percent of a pixel, then the
entire domain may be reconstructed, and then later rectified.  If the
dither steps are large, however, the fractional pixel offsets may
vary significantly over the image, requiring the reconstruction to
be done in subsets of the domain and later patched together.
This may be unattractive for some problems requiring panoramic imaging,
but may be irrelevant if the primary objects of interest are compact
or occupy only small portions of the images.

While the Fourier reconstruction method presented here works only for
translational dithers, in passing, I note that the professional image processing
literature does contain algorithms related to be present one that can combine
undersampled images with more complex geometric interrelationships.
\markcite{gran}Granrath \& Lersch (1998) present an algorithm that
constructs a Nyquist-sampled image from an image set whose members
can be related to each other with affine transformations, i. e.,
the geometric transformations that include rotation, scale change, and shear,
as well as simple translations.
The Granrath \& Lersch algorithm constructs a ``projection-onto-convex-sets''
estimate that gives the best reproduction of the image set, in
contrast to the present method, which yields a closed-form solution
to the Nyquist image.
Methods of this sort may be of interest in cases where the image
does not meet the conditions required for the present Fourier method,
but precise treatment of the Nyquist-scale is still important.

In summary, the Fourier technique presented here may not be the first
choice to construct a Nyquist image when the geometrical relationships
among the image set are complex, or the dither pattern is strongly
non-optimal.  Further, its resolution gains may appear to be superficially
modest.  Regardless, there remains a class of {\it HST} imaging problems
that push right against the diffraction scale of the instrument.
This class includes crowded field stellar photometry, the nuclear
structure of galaxies --- particularly those with bright AGN, the
morphology of lensed QSOs, and so on.
This method allows clean access to the Nyquist scale
and should be of use for these problems and more.

\acknowledgments

I wish to thank Bobby Hunt, Christoph Keller, and Ken Mighell
for useful conversations.


\clearpage

\begin{references}

\reference{fft}Bracewell, R. N. (1978), The Fourier Transform and
its Applications, p.\ 201-202 (McGraw-Hill)
\reference{cohen}Cohen, J. G., Cowie, L. L., Hogg, D. W., Songaila, A.,
Blanford, R., Hu, E. M., \& Shopbell, P. (1996), \apj, 471, L5
\reference{driz}Fruchter, A. S. \& Hook, R. N. (1998) \pasp, submitted;
astro-ph/9808087
\reference{gran}Granrath, D. \& Lersch, J. (1998), J. Opt. Soc. Am. A, 15, 791
\reference{kim}Kim, S. P., Bose, N. K., \& Valenzuela, H. M. (1990),
IEEE Transactions on Acoustics, Speech, and Signal Processing, 38, 1013
\reference{l95}Lauer, T. R., Ajhar, E. A., Byun, Y.-I., Dressler, A.,
Faber, S. M., Grillmair, C., Kormendy, J., Richstone, D., \&
Tremaine, S. 1995, \aj, 110, 2622
\reference{tsai}Tsai, R. Y. \& Huang, T. S. (1984) in
{\it Advances in Computer Vision and Image Processing,} 1, 317
(JAI:Greenwich)
\reference{hdf}Williams, R. E. et al.\ (1996), \aj, 112, 1335

\end{references}
\end{document}